\documentstyle[preprint,eqsecnum,aps]{revtex}
\begin{document}
\preprint{}
\draft

\newcommand{\bd}{\begin{displaymath}}
\newcommand{\ed}{\end{displaymath}}
\newcommand{\be}{\begin{equation}}
\newcommand{\ee}{\end{equation}}
\newcommand{\ba}{\begin{array}}
\newcommand{\ea}{\end{array}}

\title{Formation of Space-Time Structure in a Forest-Fire Model}
\author{B. Drossel and F. Schwabl}
\address{Institut f\"ur Theoretische Physik, \\
         Physik-Department der Technischen Universit\"at M\"unchen, \\
         James-Franck-Str., D-85747 Garching, Germany}
\date{\today}
\maketitle
\begin{abstract}

\end{abstract}

\section{Introduction}
\label{u1}

In nature, there exist many systems where some kind of activity propagates
without being damped. They are called excitable systems and comprise such
different phenomena as spreading of deseases, oscillating chemical reactions,
propagation of electrical activity in neurons or heart muscles, and many more
(For a review on excitable systems see e.g. \cite{tys,mer}).
All these systems exist essentially in three states which can be called
quiescent, excited, and refractory. Excitation spreads from one place to its
neighbors if they are quiescent. After excitation, a site needs some time
to recover its quiescent state. In many of these systems spiral waves
are observed. So far, these systems are mainly described by differential
equations or deterministic computer models. In this paper, we present a
stochastic forest-fire model which can be viewed as computer model for
excitable systems.
 It is defined as follows:
Each site of a $d$-dimensional hypercubic lattice of size $L^d$
is occupied by
a tree, a burning tree, or it is empty. During one timestep, the system is
parallely updated according to the following rules
\begin{itemize}
\item burning tree $\longrightarrow$ empty site
\item tree $\longrightarrow$ burning tree with
probability $1-g$  if at least one nearest neighbor is
burning
\item tree $\longrightarrow$ burning tree with
probability
$f\ll 1$  if no neighbor is burning
\item empty site $\longrightarrow$ tree with probability
$p$.
\end{itemize}

In its original version, which has been introduced by P. Bak, K. Chen, and C.
Tang, the forest-fire model contained only the tree growth parameter $p$
\cite{bak1}.

Starting with arbitrary initial conditions, this system
approaches after a short transition period a steady state the
properties of which depend on the parameter values. Throughout this paper, we
assume that the system size $L$ is large enough that no finite-size effects
occur.
In the simulations, we always chose periodic boundary conditions. Let $\rho_e$,
$\rho_t$, and $\rho_f$ be the mean density of empty sites, of trees,
and of burning trees in the steady state. These densities are
related by the equations
\be
\rho_e+\rho_t+\rho_f=1 \label{eq1}
\ee
and
\be
\rho_f=p\rho_e. \label{eq2}
\ee
The second equation says that the mean number of growing trees
equals the mean number of burning trees in the steady state.

The most interesting behavior in this model occurs for those
regions in the parameter space where the fire density decreases
to zero. Here a phase transition takes place from a steady state with fire to a
steady state without fire, and large scale structures in space and time occur.
 Eq.\ (\ref{eq2}) indicates that the fire density
decreases to zero when $p$ approaches zero. Large-scale structures can only be
expected if additionally $f\ll p$. Otherwise trees cannot live long enough to
become part
of large forests. In the case $f=0$,
the fire density  decreases to zero not only for $p\to 0$ but also when the
immunity $g$ approaches a critical value $g_C(p)$.

In this paper, we first present the
mean-field theory of the general forest-fire model which gives a rough idea of
the phase diagram. Then, we focus on the
three parameter regions indicated above: In Sect.\ \ref{u2}, we describe
quasideterministic behavior which arises in the limit $p\to 0$ with $f=g=0$. In
Sect.\ \ref{u3}, we report self-organized
critical behavior which arises for ($g=0$, $ p\to 0$, $ f/p \to 0$)
under the condition that the time scales of tree growth and burning
down of forest clusters are separated. In Sect.\ \ref{u4}, we
investigate
the percolation-like transition which takes place for $g\to
g_C(p)$. In the last section, we discuss and summarize the results.

\section{Mean-field theory}
\label{u12}

The mean-field theory is only a rough approximation of the true behavior of a
model, since spatial and temporal correlations are neglected. In mean-field
theory, the dynamics of the forest-fire model is completely described in terms
of the three densities $\rho_e$, $\rho_t$, and $\rho_f$. Each site is assumed
to be in state $i$ with the probability $\rho_i$ independently of the state of
the neighboring sites. Consequently, the change of the densities during one
timestep is given by
\begin{eqnarray}
\Delta \rho_e & = & \rho_f-p\rho_e \nonumber \\
           \Delta \rho_t & = & p\rho_e-\rho_t(1-g)
\Bigl(f+(1-f)\bigl(1-(1-\rho_f)^{2d}\bigr)\Bigr) \label{eqm1} \\
           \Delta \rho_f & = & -\rho_f+\rho_t(1-g)
           \Bigl(f+(1-f)\bigl(1-(1-\rho_f)^{2d}\bigr)\Bigr).\nonumber
\end{eqnarray}
The first equation says, that all burning trees become empty sites and that the
portion $p$ of all empty sites become populated by trees. The forest density
changes when trees
grow or catch fire. This is expressed in the second equation, where
$(1-(1-\rho_f)^{2d})$ is the probability that a tree has at least one burning
neighbor. The third equation results from the other two and equation
(\ref{eq1}). Without the lightning parameter $f$, these equations are also
derived in \cite{dro3}.

Starting with arbitrary initial conditions, this model evolves to a steady
state where the three densities are constant in time. The $\Delta \rho_i$ are
zero in the steady state. Eqs.\ (\ref{eqm1}) together with Eqs.\ (\ref{eq1})
and (\ref{eq2}) then lead to an equation for the fire density
\be \rho_f=(1-g)\bigl(1-\rho_f(1+{1\over
p})\bigr)\bigl(1-(1-f)(1-\rho_f)^{2d}\bigr).\label{eqm2}
\ee
{}From this equation, we can deduce the phase diagram of the system. A steady
state without fire, i.e. a solution of (\ref{eqm2}) with $\rho_f=0$ exists only
if $p=0$ or $g=1$ or $f=0$. When $p\neq 0$ and $g=1$ or $f=0$, the steady state
is a completely
green forest. When $p=0$, any state with $\rho_f=0$ is stationary. In the rest
of the parameter space, Eq.\
(\ref{eqm2}) has just one solution $\rho_f\neq 0$. In the plane $f=0$, Eq.\
(\ref{eqm2}) also has a solution $\rho_f\neq 0$ as long as $g<g_C$ with the
critical immunity
\be g_C=1-1/2d. \label{eqm3}
\ee
In this region, there exists a phase without fire and a phase with fire. When
the immunity $g$ approaches its critical value, the fire density in the second
phase decreases to zero. At $g_C$, a continuous transition takes place
from a region with two phases to a region with a single phase. The phase
diagram is shown in Fig.\ \ref{fig0}. Only the planes $p=0$, $g=1$ and $f=0$
are visible. In the rest of the phase space, there is a state with nonvanishing
fire density.

In this paper, we focus on phase transitions in regions where either $f=0$ or
$f/p$ and $p$ very small. Near the critical point $ f=p=0$, the fire density is
\be
\rho_f=p\Bigl(1-{1\over 2d(1-g)}\Bigr)+O(p^2)+{f\over
2d\bigl(2d(1-g)-1\bigr)}
 \bigl(1+O(p)\bigr)+O(f^2), \label{eqm4}
\ee
and near the critical point $g=g_C$, $f=0$, it is
\be
\rho_f=2d(g_C-g)/(d+1/p+1/2). \label{eqm4a}
\ee
The density of trees is at both critical points
\be
\rho_t=1-\rho_f-\rho_f/p=1/(2d(1-g)),\label{eqm5}
\ee
and a burning tree ignites a given neighbor with probability
$(1-g)\rho_t=1/2d$. The fire therefore propagates in a forest where the density
of burnable trees is at the percolation threshold, i.e. the critical
points of the mean field forest-fire model are identical to a critical point in
percolation theory.

Since the mean-field theory neglects correlations, the true behavior of the
forest-fire model is much more complicated. The mean-field theory cannot
distinguish between the limit $(\lim_{p\to 0}\lim_{f \to 0})$ where the model
shows quasideterministic behavior and the limit
$(\lim_{f/p \to 0})$ with $(p \ll f/p)$ where the model is self-organized
critical.
Another shortcoming of the mean-field theory is that it gives no dependence of
$g_C$ on $p$.

\section{The quasideterministic state}
\label{u2}

In this section, we consider the case $f=g=0$ with $p$ very small. Fire spreads
from burning trees to their neighbors, but cannot
occur spontaneously. This is the forest-fire model as originally
introduced by P. Bak, K. Chen, and C. Tang \cite{bak1}, which is -- in contrast
to earlier assumptions -- not self-organized critical \cite{gra1,mos}.
When the initial conditions are chosen in such a way that
the fire does not die during the first few timesteps,
the system develops a steady state with nonvanishing fire density. With
decreasing $p$, the fire fronts assume a more and more regular, spiral-shaped
form \cite{gra1}. A snapshot of
such a steady state is shown in Fig.\ \ref{fig1}. A quantitative analysis of
the spatial and temporal correlations of the fire reveals that the system has a
characteristic length scale and a characteristic time scale both of which are
proportional to $1/p$ and become more distinct with decreasing $p$ \cite{mos}.
The model therefore becomes more and more deterministic with decreasing $p$.
The origin of the spirals is the following:
Fire fronts separate empty areas from forest areas. They move into the forest,
i.e. the empty area grows at the cost of the forest. At the end of a fire
front, there is a motion in the opposite direction: since there is no fire, the
forest grows into the empty area. So the fire front and the tree growth
``front''
wind around each other, and the  end of the fire front becomes a spiral center.
The distance between two windings of a spiral is of the order $1/p$ which
is the time the forest needs to grow again after it has been burned. This
is the origin of the time and length scales mentioned above.
There even is an explanation for the increasing determinism for small $p$. When
$p$ is decreased by a factor $b$, the length and time scale in the system are
increased by the same factor. Between two spiral arms, there are $b$ times as
many
sites than before. On a length scale proportional to the inverse tree growth
rate, the forest density
in front of the fire varies less than before (since the average is taken over
more sites), and the fire front therefore
becomes increasingly smooth.

\section{The self-organized critical state}
\label{u3}

In this section again we set $g=0$ and introduce the lightning probability $f$.
The ratio $f/p$ determines how many trees grow between two lightnings. We
consider the case where $f/p$ is small and choose for a given value of $f/p$ a
tree growth rate $p$ which is so small that a forest cluster which is struck by
lightning burns down before new trees grow at its edge. Under this condition
large and small forest
clusters burn down in the same way, i.e. large and small fires are similiar.
In this situation,
there occur fires of any size in the system, since lightning might strike any
forest cluster and since there exist very large forest clusters when $f/p$ is
small. So the system is self-organized critical \cite{dro1} in the sense of
Bak, Tang, and Wiesenfeld \cite{bak2}. Fig.\ \ref{fig2}
shows a snapshot of the steady state. The double time scale
separation (tree growth occurs much more often than lightning, and burning down
of forest clusters is much faster than tree growth) is the essential condition
required for self-organized critical behavior in the forest-fire model.
The system is
self-organized because the steady state is independent of the initial
conditions and independent of the exact values of the parameters as long as
time scales are separated in the manner mentioned above.
 It is critical because there
are power-law  correlations over long distances and long time intervals.

In the following, we first present a scaling theory for the self-organized
critical state (subsection \ref{u30}). Then we show results of computer
simulations in two dimensions which confirm this scaling theory (subsection
\ref{u31}). In one dimension, the self-organized critical forest-fire model is
nontrivial, and yet exact results can be obtained. This is shown in subsection
\ref{u32}.

\subsection{Scaling theory}
\label{u30}

Let $\bar s$ be the mean number of trees
destroyed by a lightning stroke. In the steady state it equals
the mean number of trees growing between two lightning strokes
and is given by
\be
\bar s=p(1-\rho_t)/f\rho_t. \label{eq3}
\ee
In the limit $f/p\to 0$, this number diverges proportionally to
$(f/p)^{-1}$, since $\rho_t$ must approach a constant value for
small $f/p$.
Eq.\ (\ref{eq3}) represents a power law indicating a critical point
in the limit $f/p \to 0$.

 Let $n(s)$ be the mean number of
forest clusters per unit volume consisting of $s$ trees.
Then the mean forest density is
\be
\rho_t =\sum_1^\infty sn(s), \label{eq4}
\ee
and the mean number of trees destroyed by a lightning stroke is
\be
\bar s = \sum_1^\infty s^2 n(s) /\rho_t. \label{eq5}
\ee
Since $\rho_t(f/p\to 0)$ is finite and $\bar s$ diverges $\propto
(f/p)^{-1}$, these equations imply that $n(s)$ decreases at least
like
$s^{-2}$ but not faster than $s^{-3}$. As long as the system is not
exactly at the critical point $f/p=0$, i.e. for finite $f/p$,  there must be a
cutoff in the cluster size distribution for very large forest clusters.
 We conclude that \cite{dro1}
\be
n(s)\propto s^{-\tau}{\cal C}(s/s_{\text{max}}) \label{eq6}
\ee
with $2\le \tau \le 3$ and
\be
 s_{\text{max}}(f/p)\propto (f/p)^{-
\lambda}\propto \bar s^\lambda.\label{eq7}
\ee
The cutoff function ${\cal C}(x)$ is more or less constant for
$x\le 1$ and
decreases to zero for large $x$. Eqs.\ (\ref{eq5}) -- (\ref{eq7}) yield $\bar s
\propto s_{\text{max}}^{3-\tau}$, which leads to the scaling relation
\be
\lambda=1/(3-\tau).\label{eq8}
\ee

In the case $\tau=2$, the right-hand side of Eq.\ (\ref{eq6}) acquires a
factor $1/\ln(s_{\text{max}})$ since the forest density given by Eq.\
(\ref{eq4})
must not diverge in the limit $f/p\to 0$. The
mean number of forest clusters per unit volume $\sum_1^\infty
n(s)$, therefore, decreases to zero for $f/p\to 0$, and consequently
the forest density approaches the value 1.

We also introduce the cluster radius $R(s)$ which is the mean distance of the
cluster trees from their center of mass. It is related to the cluster
size $s$ by
\be
s\propto R(s)^\mu \label{eq9}
\ee
with the fractal dimension $\mu$. The correlation length $\xi$ is defined by
\begin{eqnarray*}
\xi^2 &=& \sum_{s=1}^\infty s^2n(s)R^2(s) / \sum_{s=1}^\infty s^2 n(s)\\
&\propto & (f/p) \int_1^\infty s^{2-\tau+2/\mu} {\cal C}(s/s_{\text{max}})
\text{d} s \propto (f/p) s_{\text{max}}^{3-\tau+2/\mu} \int_0^\infty
x^{2-\tau+2/\mu} {\cal C}(x) \text{d} x \\
& \propto & (f/p)^{-2\lambda/\mu}.
\end{eqnarray*}
We conclude
\be
\xi \propto (f/p)^{-\nu} \text{ with }  \nu = \lambda / \mu. \label{eq10}
\ee

Another quantity of interest is the mean cluster radius
\bd
\bar R = \sum_{s=1}^\infty s n(s) R(s) / \sum_{s=1}^\infty sn(s)
\, \left\{ \propto
(f/p)^{-(\nu-\lambda+1)}\hbox{ , if } \nu -\lambda +1 \ge 0 \, ;
\atop  = \hbox { const., if } \nu-\lambda+1 < 0 \, .\hfill
\right.
\ed
This leads to
\be
\bar R \propto (f/p)^{-\tilde \nu} \text{ with }
\tilde\nu=\text{Min}(0,\nu-(\lambda-1))\, .
\label{eq9a}
\ee

The mean forest density $\rho_t$ approaches its critical value $\rho_t^c =
\lim_{f/p \to 0} \rho_t$ via a power law
\be
\rho_t^c - \rho_t  \propto (f/p)^{1/\delta}. \label{eq9aa}
\ee

Finally, we introduce some exponents characterizing the temporal behavior of
the fire.
Let $T(s)$ be the average time a cluster of size $s$ needs to burn down when
ignited, and
$N(T)$ the portion of fires that live exactly for $T$ timesteps. Then the
exponents $b$ and $\mu'$ are defined by
\be s\propto (T(s))^{\mu'} \text{ and } N(T) \propto T^{-b}. \label{eq9b}
\ee
{}From
\bd
N(T) \text{d} T \propto sn(s) \text{d} s
\ed
follows the scaling relation
\be
b=\mu'(\tau-2) +1. \label{eq9c}
\ee
The timescale of the system is set by
\be
T_{\text{max}}=T(s_{\text{max}}) \propto (f/p)^{-\nu'}
\label{eq9c1}
\ee
with
\be
\nu' = \lambda/\mu'. \label{eq9c2}
\ee
The dynamical critical exponent $z$ is defined by
\bd
T_{\text{max}} \propto \xi^z,
\ed
which leads to
\be
z=\mu/\mu'=\nu'/\nu. \label{eq9f}
\ee
The condition of time scale separation now can be expressed in terms of the
critical exponents and reads
\be
(f/p)^{-\nu'}\ll p^{-1} \ll f^{-1},
\ee
or equivalently
\be
f\ll p \ll f^{\nu'/(1+\nu')}.
\ee
The average lifetime of fires is
\be
\bar T = \sum_{s=1}^\infty sn(s) T(s) /\sum_{s=1}^\infty sn(s) \propto
(f/p)^{-\tilde \nu'}
\ee
with
\be
\tilde \nu' = \text{Min}(0,\nu' - (\lambda-1))\, .
\ee
The average number $N_s(t)$ of trees that burn $t$ timesteps after a cluster of
size $s$ is struck by lightning enters the definition of
 the temporal fire-fire correlation function $G(\tau)$
\bd
G(\tau)  \propto \sum_{s=1}^\infty n(s) s \sum_{t=0}^{\infty}
 N_s(t) N_s(t+\tau).
\ed
The power spectrum is the Fourier transform of the fire-fire correlation
function
\be
G(\omega) = 2\int_0^\infty G(\tau) \cos(\omega \tau) \text(d) \tau \propto
\omega^{-\alpha} \text{ for small $\omega$. }\label{eq9e}
\ee

\subsection{Critical exponents in two dimensions}
\label{u31}

In two dimensions, the values of the critical exponents were determined by
computer simulations
\cite{cla} in systems of size $\le 16384^2$ and with parameters $f/p \ge
1/32000$. Fig.\ \ref{fig3} shows
the distribution $sn(s)$ as function of $s$ which gives the critical exponent
$\tau$. The values of all critical exponents  are given in Table~\ref{tab1}.
Except for $z$ and $b$, these exponents have directly been determined in
simulations.  They satisfy the scaling relations
derived in subsection \ref{u30}.
The exponents $\tau$, $\nu$, $\delta$ and $\lambda$ have also been determined
by P. Grassberger \cite{gra2}. The
values of $\tau$, $\nu$, and $\delta$ agree with ours, but not the value of
$\lambda$. Our result for $\lambda$  satisfies the above scaling
relations (\ref{eq8}), (\ref{eq10}), and (\ref{eq9a}).
The value of the critical forest density is $\rho_t^c=0.4081(7)$ in agreement
with \cite{gra2}.
In \cite{dro1} simulations on samples of size $500^2$ seemed to suggest
classical values of the critical exponents ($\tau=\mu=2$).

\subsection{Critical exponents in one dimension}
\label{u32}

Even in one dimension, the self-organized critical forest-fire model is
nontrivial.
Since there exist no infinitely large forest clusters for $\rho_t < 1$, the
forest density has to approach the value 1 in the critical limit $f/p \to 0$.
{} From the consideration
which follows Eq.\ (\ref{eq8}) we conclude $\tau=2$. The value of this and
other critical exponents as well as other  exact results can be determined
analytically. A
detailed calculation is given in \cite{dro2}. A more intuitive approach is
presented in the following.
Consider a string of $n$ sites with $n \ll s_{\text{max}}$. The probability
that lightning strikes this string is very small. Therefore trees grow on it
until it is covered by a completely dense forest.  Just before this dense
forest portion  burns down, it is part of a large forest which will be struck
by lightning somewhere. After the fire has passed our string, it will be
completely empty. Let $P_n(m)$ be the probability that the string is covered by
$m$ trees. Since trees grow randomly and since fire does not pass the string
before all its trees are grown, the $P_n(m)$ are related by
\begin{eqnarray*}
p n P_n(0) & = & p(1-\rho_t), \\
p(n-m)P_n(m) & = & p(n-m+1)P_n(m-1)
\text{ for } m \neq 0,n,
\end{eqnarray*}
which lead to the result
\begin{eqnarray}
P_n(m) & = & (1-\rho_t)/(n-m) \text{ for } m<n, \nonumber \\
P_n(n) & = & 1-(1-\rho_t)\sum_{m=0}^{n-1}1/(n-m) \label{eq11} \\
& = & 1-(1-\rho_t) \sum_{m=1}^n 1/m. \nonumber
\end{eqnarray}
Eq.\ (\ref{eq11}) is exact up to terms of order $f/p$ and thus the more
accurate the closer to the critical point.
The size distribution of forest clusters is
\begin{eqnarray}
n(s) & = & P_{s+2}(01\dots 10)={P_{s+2}(s)\over
{s+2 \choose s}} \nonumber \\
& = & {1-\rho_t\over (s+1)(s+2)} \simeq (1-\rho_t)s^{-2}.
\label{eq12}
\end{eqnarray}
This is a power law with the critical exponent $\tau=2$. Fig.\ \ref{fig4} shows
the size distribution
$sn(s)$ for fires as determined by computer simulations for $f/p=1/25000$. The
smooth curve is
the analytic result (\ref{eq12}) which fits the simulation perfectly as long as
$s < s_{\text{max}}\simeq 1000$.
The size distribution $n_e(s)$ of clusters of empty sites is
\begin{equation}
n_e(s)=P_{s + 2}(10 \dots 01) = {P_{s + 2}(2) \over
{s + 2 \choose 2}} = {2 (1 - \rho_t) \over s (s + 1) (s + 2)},
\label{eq12a}
\end{equation}
which is also derived in \cite{pac93}.
The result (\ref{eq11}) contains much more information beyond the cluster size
distribution. So it can be shown that a mean number of $s+2$ trees is added to
a forest cluster of size $s$ when a tree grows at its edge, i.e. the growth
speed of forest clusters is proportional to their size.

In the limit $f/p \to 0$, $s_{\text{max}}$ diverges and $n$ may become
large. Nevertheless $P_n$ must remain $\ge 0$.
It follows
\be
(1-\rho_t) \propto [\ln(s_{\text{max}})]^{-1},
\ee
 i.e. the forest
density approaches logarithmically the value one in the
limit $f/p
\to 0$, and the mean number of forest clusters per unit length
decreases to zero.

We calculate
$s_{\text{max}}$ from the
condition that a string of size $n
\le s_{\text{max}}$ is not
struck by lightning until all trees are
grown. When a string of size $n$ is completely empty at time
$t=0$, it will be occupied by $n$ trees after
\be
T(n)=(1/p)\sum_{m=1}^n 1/m \simeq \ln(n)/p \label{eq12b}
\ee
 timesteps on an
average. The mean number of trees after $t$ timesteps follows from the growth
equation
\bd
\dot \rho_t = p(1-\rho_t)
\ed
and is
\bd
m(t)=n[1-
\exp(-pt)].
\ed
 The probability that lightning
strikes a string of size $n$ before all trees are grown is
\bd f\sum_{t=1}^{T(n)} m(t)\simeq (f/p)n(\ln(n)-1) \simeq
(f/p)n\ln(n)
\ed
 We conclude
\begin{equation}
s_{\text{max}}\ln(s_{\text{max}})
\propto p/f \; \text{for large $p/f$},
\end{equation}
and, to leading order in $\ln(f/p)$
\be
s_{\text{max}} \propto {p\over f}{1\over \ln(p/f)}.
\ee
Consequently the critical exponent $\lambda$ is $\lambda=1$. The remaining
critical exponents can directly be derived using scaling relations or the
defining equations  given in subsection \ref{u30}. They are given in
Table~\ref{tab1}.

\section{The percolation phase transition}
\label{u4}

We now consider the case $f=0$ and $g>0$.
Watching a 2-dimensional system on a video screen, one realizes
that the fire
fronts present for $g=0$ become more and more fuzzy with
increasing immunity $g$ and that the forest grows denser.

At the critical value $g=g_C(p)$, the fire density goes to zero, and
the critical fire spreading resembles percolation.  For $g>g_C$,
the steady state of the system is a completely green forest. Fig.\ \ref{fig5}
shows a
snapshot of the system for a value of $g$ just below $g_C$.
Fig.\ \ref{fig5a} shows the
 density of empty sites $\rho_e=\rho_f/p$ as function of $g$ for
$p=0.1$ as determined from computer simulations. The fire density
depends linearly on $(g_C-g)$ near the critical line. This linear
dependence can already be obtained from mean-field theory  \cite{dro3}.
In the general case it can be understood by the following argument.
Consider the stationary state near $g_C$. Since $\rho_f=p\rho_e$,
there exist for each burning tree $p^{-1}$ empty
sites on an average. Most of these empty sites have been visited
by the fire less than $ \simeq 1/p$ timesteps ago and
are situated within a distance $\le 1/p$ of the
burning trees. The fire spreading therefore is influenced to a
great deal by empty sites (and even other burning sites). Let
$3-\alpha(g)$ be the mean number of trees
that are nearest neighbors of burning trees, divided by the
number of burning trees. Obviously $\alpha(g)>0$, since each fire
site has at least one empty neighbor namely the site where the
fire has been one timestep before. The larger $\alpha(g)$,
the more is the fire spreading hindered by empty sites and other
burning sites. The mean number
of
trees that are ignited by a burning tree is $(1-g)
(3-\alpha(g))$. In the stationary state
this has to be one tree on an average, i.e.
\bd
\alpha(g)=3-1/(1-g).
\ed
In mean-field theory (with $2d$ replaced by $(2d-1)$) $g_C=2/3$ and
$\alpha(g_C)=0$.  The fact that actually  $\alpha(g_C)> 0$ for all values of
$p$ (since $g_C<2/3$) indicates that a fire is
influenced by the presence of other fires even in the
limit of zero fire density. A
burning tree often ignites two or three trees even at $g=g_C$,
and these fires interfere with each other when they spread
further.
We now show that $(\partial \rho_f(g)/\partial
g)_{g=g_C}$ is finite and nonvanishing,
i.e. that any change in $g$ is
accompanied by a change in $\rho_f$ and vice versa. To this end we
first consider the system exactly at the critical line. Then the
fire
density is zero, i.e. an infinitesimally small portion of all
sites in
an infinite system are burning sites. The fire spreads by a
branching-and-death mechanism: A burning tree may ignite two or
three neighbors (a branching process), one neighbor, or no
neighbor (it dies). These processes are in an equilibrium in the
steady state since the fire density is constant in time. On an
average a burning tree produces just another burning tree. We
mentioned already above that these branching-and death processes
always are influenced by other burning sites since $\alpha(g_C)>
0$. This is not only true for  single burning sites.
Groups of burning sites also interfere with each other.
Otherwise it would be possible to fill the system with these
groups up to a finite fire density - in contradiction with the
observation that the fire density is zero. Now we
decrease the immunity slightly from its critical value. Since
less trees are immune, more branching processes take place than
before, and the fire density increases until a new equilibrium is
established. We just have seen that there always is an
interaction between fire sites. Therefore the
newly created fire sites cannot evolve independently of the rest
of
the system, but they interfere immediately with the fire sites
that
have been in the system before the decrease of $g$. As a
consequence
 $\alpha$ increases proportionally to the fire density. There
cannot be a change in the fire density
without a corresponding change in $\alpha$. Since on the other
hand, via Eq.(11), changes of $\alpha$ are related linearly to
changes of $g$, the derivative  $(\partial
\rho_f/\partial g)_{g=g_C}$ is finite.

Fig.\ \ref{fig6} shows the critical line $g_C(p)$ as obtained
from computer simulations. The critical immunity $g_C$ increases
with increasing $p$. This is plausible since with increasing $p$
the fire can sooner return to sites where it already has been.
The main features of $g_C(p)$ can be derived by the following consideration
\cite{dro3}. If the immunity is at its critical value and if the
initial state is an insolated burning tree in an infinitely large
forest, the mean number of burning trees after a long time
interval is again a single burning tree, i.e. the fire is just
able to survive. We therefore started with an isolated burning
tree with one empty neighbor (the site where the fire has been
one timestep ago) and calculated the mean number $F(t,p,g)$ of
burning trees after $t$ timesteps. An approximate value of $g_C$
is given by the condition $F(t,p,g_C)=1$. The larger $t$, the
more accurate is the result. Fig.\ \ref{fig7} shows $g_C(p)$ for $t=4$. This
curve is linear for small $p$ and flat for $p\to 0$, as obtained
from the simulations.

Finally, we comment on the nature of the percolation phenomenon
occuring at $g_C$. From the simulations we obtained $\lim_{p\to
0} g_C(p)\simeq 0.467$. Had we introduced the immunity by
making sites or bonds between sites permanently immune, we would
have obtained a value $g_C(p=0)=0.41$ or 0.5, which is just one
minus the percolation threshold for site or bond percolation. The
critical immunity in this case were independent of $p$ since
there could not exist an infinite path for the fire through the
system for $p>0$ if it did not for $p=0$. In our model, sites are
not
permanently immune and the fire can take paths that it
cannot take in site percolation. The percolation threshold $1-
g_C(p=0)$ therefore is lower than in site percolation.  We call
the new kind of
percolation observed in this model ``fluctuating site
percolation''.

\section{Summary and discussion}
\label{u5}

In this paper, we have shown that a simple stochastic forest-fire model shows a
variety of different phenomena. Depending on the range of parameter values,
quasideterministic spiral-shaped fire fronts, self-organized critical behavior,
or percolation-like behavior is observed. An essential condition for the
occurence of self-organized critical behavior in the forest-fire model is a
double separation of time scales. When time scales are not separated,
completely different phenomena can be observed. The origin of the mentioned
phenomena was explained by a series of simple physical arguments. Computer
simulation results were presented, and analytic calculations including a
mean-field theory and a scaling theory were performed. For the one-dimensional
self-organized critical
forest-fire model exact results were derived.

The forest-fire model can be viewed as a computer model for excitable systems.
Thus, it can be
expected that the phenomena occuring in the forest-fire model should also be
observed in excitable systems, when the appropriate range of parameter values
is investigated. Spiral waves in two- or higher-dimensional excitable systems
are quite familiar, but percolation and self-organized critical behavior have
not been reported so far. We hope that this paper will stimulate research aimed
at the detection of these phenomena in excitable systems.

\subsection*{Acknowledgement}

We thank S. Clar for producing Figs.3-6 and S. Clar, W.
Mo\ss ner, and N. Knoblauch for writing the simulation codes.

\begin{table}
\caption{Critical exponents in one and two
dimensions. Quantities with  $^*$ have logarithmic corrections.
 }
\begin{tabular}{ccccccccccccc}
$d$ & $\tau$ & $\mu$ & $\lambda$ & $\nu$ & $\tilde \nu$ & $\mu'$ &  $\nu'$ &$
\tilde \nu'$ & $z$ & b &
${1/\delta} $ & $\alpha$ \\
\tableline
1 & 2 & 1 & $1^*$ & $1^*$ & $1^*$ & 1 & $1^*$ & $1^*$ & 1 & 1 & $0^* $ & $2^*$
\\
2 & 2.14(3) & 1.96(1) & 1.15(3) & 0.58 & 0.43 & 1.89(3) & 0.61(3)  & 0.44 &
1.04(1) & 1.27(7) & 0.48(2) & 1.72(5) \\
\end{tabular}
\label{tab1}
\end{table}


\begin{figure}
\caption{Phase diagram derived from mean-field theory. In the inner of the cube
is a phase with nonvanishing fire density.}
\label{fig0}
\end{figure}

\begin{figure}
\caption{Snapshot of the Bak et al forest-fire model in the steady state for
$p=0.005$ and $L=800$. Trees are black, empty sites are white. Burning trees
are too small to be seen. They are situated at the sharp black-white
interface. }
\label{fig1}
\end{figure}

\begin{figure}
\caption{Snapshot of the self-organized critical state for $f/p=1/200$ and
$L=500$. Trees are black, empty sites are white. }
\label{fig2}
\end{figure}

\begin{figure}
\caption{Size distribution $s n(s)$ of the fires for $f/p=1/16000 $ and
$L=8192$ in two dimensions.}
\label{fig3}
\end{figure}

\begin{figure}
\caption{Size distribution of the fires for $f/p=2/50000 $ and $L=2^{20} $ in
one dimension.
The smooth line is the theoretical result, which is valid for cluster
sizes $\le s_{\text{max}}$.}
\label{fig4}
\end{figure}

\begin{figure}
\caption{Snapshot of the forest-fire model near the
critical immunity for $p=0.1$, $g=0.48$, $L=200$. Trees are grey,
empty sites are white, and burning trees are black.}
\label{fig5}
\end{figure}

\begin{figure}
\caption{The density of empty sites as function of immunity for $p=0.1$ and
$L=500$. }
\label{fig5a}
\end{figure}

\begin{figure}
\caption{The critical immunity $g_C(p)$ obtained from computer simulations.}
\label{fig6}
\end{figure}

\begin{figure}
\caption{The critical immunity $g_C(p)$ as calculated for $t=4$. }
\label{fig7}
\end{figure}

\end{document}